\documentclass[final]{pasj02}
\usepackage{url}
\usepackage{bm}
\PassOptionsToPackage{pdftex}{graphicx}

\Received{$\langle$reception date$\rangle$}
\Accepted{$\langle$acception date$\rangle$}
\Published{$\langle$publication date$\rangle$}

\begin{document}

\title{ 
Inhibition of Accretion by the Stellar Wind in Misaligned Be/X-ray Binaries
}

\author{Atsuo T. \textsc{Okazaki}\altaffilmark{1}%
}
\altaffiltext{1}{Faculty of Engineering, Hokkai-Gakuen University, Sapporo 062-0911, Japan}
\email{okazaki@hgu.jp}




\maketitle

\begin{abstract}
Be/X-ray binaries (BeXRBs) constitute a major subclass of high-mass
X-ray binaries.  They show intermittent X-ray activity with
$L_X \gtrsim 10^{36}\ \mathrm{erg\,s^{-1}}$, while remaining quiescent
most of the time with $L_X \lesssim 10^{34}\ \mathrm{erg\,s^{-1}}$.
BeXRBs generally have eccentric orbits as a result of supernova kicks when neutron stars were born. In these systems, the same kicks are also likely to make the binary orbital axis misaligned with the spin axis of the Be star. In such systems, when the neutron star captures gas from the equatorial disk of the Be star, 
the resulting accretion disk is in general tilted to both the Be disk plane and to the binary orbital plane. This raises an interesting possibility that in misaligned BeXRBs, the polar wind of the Be star collides with the accretion disk and significantly affects its structure by the large ram pressure. In this paper, we study the effects of the stellar wind on the accretion dynamics in misaligned BeXRBs. Using analytical wind and disk models, we first compare the wind's ram pressure with the gas pressures of the accretion flow to derive a condition for the stellar wind to strongly suppress accretion, and then apply the condition to a sample of BeXRBs whose relevant parameters are well determined or constrained. We find that wind-driven inhibition is a plausible mechanism for
suppressing accretion in systems with slowly rotating neutron stars in
wide orbits, where the classical propeller mechanism is expected to be
inefficient.  The effect is particularly important if the accretion flow
is hot and low-density, or after the accretion rate has declined from the
outburst level.
\end{abstract}


\section{Introduction}
\label{sec:intro}

High-mass X-ray binaries offer insights into active binary interactions in an evolutionary stage prior to the formation of double neutron star binaries. 
About half of high-mass X-ray binaries in the Milky Way galaxy and the vast majority of those in the Magellanic Clouds consist of a neutron star and a Be star (e.g., \cite{Fornasini2023}, and references therein). Here, a Be star is a massive star with a fast polar wind driven by stellar radiation and a dense equatorial disk formed by viscous diffusion of gas ejected from the stellar equatorial surface (e.g., \cite{Rivinius2013}). 
These Be/X-ray binaries (BeXRBs) are transient X-ray sources. They are quiescent most of the time, but occasionally exhibit two types of X-ray outbursts: normal (or Type I) outbursts, of which the X-ray luminosity is moderate in the range $L_\mathrm{X} \sim 10^{36-37}\,\mathrm{erg\;s}^{-1}$, and giant (or Type II) outbursts, which are significantly brighter ($L_\mathrm{X} > 10^{37}\,\mathrm{erg\;s}^{-1}$) and less frequent than normal outbursts \citep{Stella1986, Negueruela1998}. 

The transition between the outburst and quiescent states has been witnessed in a few systems with relatively short spin periods ($P_\mathrm{spin} \sim$ a few seconds), such as 4U~0115$+$63, V~0332$+$53 \citep{Stella1986, Campana2001, Tsygankov2016}, and SMC~X-2 \citep{Lutovinov2017}. In these systems, the X-ray luminosity rapidly increased/decreased by almost two orders of magnitude between $\sim 10^{34}\;\mathrm{erg\;s}^{-1}$ and $\sim 10^{36}\;\mathrm{erg\;s}^{-1}$. Such a rapid change of X-ray luminosity as well as softening of the X-ray spectra in the low state strongly indicate that the centrifugal inhibition of accretion by the rapidly rotating magnetosphere of the neutron star (or propeller mechanism) is responsible for the quiescent state, stopping the accretion flow \citep{IllarionovSunyaev1975}.

Given that the spin period of the neutron star ranges over almost three orders of magnitude in BeXRBs, however, it is unlikely that the propeller mechanism is the sole agent for state transition in all BeXRBs. In this paper, we propose that the stellar wind from the Be star is another mechanism that causes the state transition in some systems, particularly in those hosting neutron stars with long spin periods.

Since the BeXRBs have, in general, eccentric orbits as a result of non-spherically symmetric supernova explosion when a neutron star was born (e.g., \cite{Knigge2011}, and references therein), it is likely that the binary orbital plane in these systems is misaligned with the equatorial plane of the Be star. In such systems, when the neutron star captures gas from the Be star's disk,
the resulting accretion disk is most likely tilted to the equatorial plane of the Be star as well. This leads to an interesting possibility that in misaligned BeXRBs, the stellar wind of the Be star collides with the accretion flow and significantly affects its structure by the large ram pressure, completely dispersing accreting gas in an extreme situation. 

In this paper, after describing our model in section~\ref{sec:models}, we formulate an analytic criterion for wind inhibition of accretion
in section~\ref{sec:criterion} by comparing the wind ram pressure with the characteristic
pressure of the accretion flow. In section~\ref{sec:indiv} we apply this criterion to a
sample of BeXRBs with well determined or constrained stellar, orbital,
and neutron-star parameters, thereby providing a systematic framework for
assessing the relative importance of wind-driven ablation and the
classical propeller mechanism in producing X-ray quiescence. Section~\ref{sec:discussion}
discusses the physical implications of our results, the limitations of
the instantaneous ablation criterion, and the connection with previous
hydrodynamical simulations. The paper is concluded in section~\ref{sec:conclusions}.

\section{Models for accretion flows and stellar winds}
\label{sec:models}

\subsection{Standard accretion disk}
\label{sec:sd}

Using a one-zone approximation, the Shakura-Sunyaev disk solution \citep{ShakuraSunyaev1973} in an outer region, where the density and the temperature are low and the opacity is dominated by the free-free absorption, is given by
\begin{eqnarray}
   &&\Sigma = 1.7 \alpha_{0.1}^{-4/5} \dot{m}_{16}^{7/10} m_{1.4}^{1/4} r_{12}^{-3/4}
      \;\mathrm{g\;cm}^{-2},
      \label{eq:sigma_sd} \\
   &&H = 2.4 \times 10^{10} \alpha_{0.1}^{-1/10} \dot{m}_{16}^{3/20} m_{1.4}^{-3/8}
      r_{12}^{9/8}\;\mathrm{cm},
      \label{eq:h_sd} \\
   &&\rho = 3.6 \times 10^{-11} \alpha_{0.1}^{-7/10} \dot{m}_{16}^{11/20} m_{1.4}^{5/8} 
      r_{12}^{-15/8}\;\mathrm{g\;cm}^{-3},
      \label{eq:rho_sd} \\
   &&T_\mathrm{c} = 6.5 \times 10^2 \alpha_{0.1}^{-1/5} \dot{m}_{16}^{3/10} m_{1.4}^{1/4} 
      r_{12}^{-3/4}\;\mathrm{K}, 
      \label{eq:t_sd} \\
   &&P_\mathrm{gas} = 3.1 \alpha_{0.1}^{-9/10} \dot{m}_{16}^{17/20} m_{1.4}^{7/8} r_{12}^{-21/8}
   \;\mathrm{g\;cm}^{-1}\mathrm{s}^{-2}
   \label{eq:pgas_sd}
   \label{eq:pgas_out}
\end{eqnarray}
(e.g., \cite{Frank2002,Kato2008}), 
where $\Sigma$ is the surface density, 
$H$ is the vertical scale-height of the accretion disk,
$\rho$ is the vertically averaged density defined as $\Sigma = 2 \rho H$,
$T_\mathrm{c}$ is the mid-plane temperature,
and $P_\mathrm{gas}$ is the gas pressure given by $P_\mathrm{gas} = \rho k T_\mathrm{c}/(\mu m_\mathrm{H})$,
where $k$ is the Boltzmann constant, 
$\mu = 0.62$ is the mean molecular weight for a fully ionized plasma with cosmic abundances, 
and $m_\mathrm{H}$ is the mass of a hydrogen atom.
On the right-hand-side of these equations, 
$\alpha_{0.1}$ is the Shakura-Sunyaev's viscosity parameter normalized by 0.1, 
$r_{12}$ is the distance from the neutron star normalized by $10^{12}$\;cm, 
$\dot{m}_{16}$ is the accretion rate normalized by $10^{16} \mathrm{g\;s}^{-1}$ ($\simeq 1.6 \times 10^{-10} M_\odot\;\mathrm{yr}^{-1}$),
and $m_{1.4}$ is the mass of the neutron star normalized by $1.4\;M_\odot$.

\subsection{Self-similar ADAF}
\label{sec:adaf}

Unlike standard disks, where the viscous heating is locally balanced with the radiative cooling, the advection cooling cannot be neglected in hot accretion flows. It is particularly important for low accretion rates $\dot{m} \ll 1$ (or $\dot{m}_{16} \ll 20 m_{1.4}$), where the radiative cooling is inefficient.
\citet{Narayan-Yi1994} derived a set of solutions of 
self-similar, optically thin, advection dominated accretion flows (ADAFs).
Therefore, in this study we consider these flows as another model for accretion flows onto the neutron star.
By adopting the self-similar scaling law proposed by \citet{Narayan-Yi1994} and \citet{Narayan-Yi1995},
we obtain a set of solutions as 
\begin{eqnarray}
   &&v_r = -1.4 \times 10^6
     c_1\alpha_{0.1} m_{1.4}^{1/2} r_{12}^{-1/2}\;\mathrm{cm\;s}^{-1}, 
     \label{eq:vr_adaf} \\
   &&c_\mathrm{s} = 1.4 \times 10^7
     c_3^{1/2} m_{1.4}^{1/2} r_{12}^{-1/2}\;\mathrm{cm\;s}^{-1},
     \label{eq:cs_adaf} \\
   &&H = 1.0 \times 10^{12}c_3^{1/2} r_{12}\;\mathrm{cm},
     \label{eq:h_adaf} \\
   &&\rho = 5.8 \times 10^{-16}
     c_1^{-1}c_3^{-1/2}\alpha_{0.1}^{-1} \dot{m}_{16} m_{1.4}^{-1/2}r_{12}^{-3/2} \nonumber\\
   && \hspace{0.15\hsize} \mathrm{g\;cm}^{-3},
     \label{eq:rho_adaf} \\
   &&P_\mathrm{gas} = 1.1 \times 10^{-1} c_1^{-1} c_3^{1/2} \alpha_{0.1}^{-1} \dot{m}_{16} m_{1.4}^{1/2} r_{12}^{-5/2} \nonumber\\
   && \hspace{0.15\hsize} \mathrm{g\;cm}^{-1}\mathrm{s}^{-2},
     \label{eq:pgas_adaf} \\
   &&P_\mathrm{ram} = 1.1 \times 10^{-3}  c_1 c_3^{-1/2} \alpha_{0.1} \dot{m}_{16} m_{1.4}^{1/2} r_{12}^{-5/2} \nonumber\\
   && \hspace{0.15\hsize} \mathrm{g\;cm}^{-1}\mathrm{s}^{-2}.
 \label{eq:pram_adaf}
\end{eqnarray}
where  $v_{\rm{r}}$ is the radial velocity, 
$c_\mathrm{s}$ is the isothermal sound speed,
$\rho$ is the density calculated by $\rho = \dot{M}/(4 \pi r H |v_r|)$,
$P_\mathrm{gas} = \rho c_\mathrm{s}^2$ and $P_\mathrm{ram} = \rho v_r^2$ are the gas and ram pressures, respectively, and $c_1$ and $c_3$ are constants. For simplicity, we assume that all the viscously dissipated energy is advected and that $\alpha \ll 1$ and the specific heat ratio is 3/2, for which $c_1$ and $c_3$ are given as $c_1 = 0.53$ and $c_3 = 0.35$.
Note that $P_\mathrm{gas} \gg P_\mathrm{ram}$ always holds for $\alpha^2 \ll 1$.

\subsection{Stellar wind}
\label{sec:wind}

We approximate the Be-star wind near the neutron star by a spherical
wind with a terminal velocity $v_\mathrm{sw}$. This is justified
because, for most BeXRBs considered here, the periastron separation is
much larger than the stellar radius, so that the wind speed near the
neutron star is already close to the terminal value.\footnote[1]{We have checked that using a beta-law velocity field instead of the
terminal-speed approximation changes the resulting critical accretion
rates only weakly for the present sample.}
We adopt 
$v_\mathrm{sw} = 2.6 v_\mathrm{esc}$ \citep{Vink2001} with $v_\mathrm{esc} = (2G M_{*}/R_{*})^{1/2}$ being the escape velocity, where $M_{*}$ is the stellar mass.
Regarding the wind mass-loss rate $\dot{M}_\mathrm{sw}$, we calculate it by using the mass-loss rate recipe by \citet{Bjorklund2023}. 
In this model, the wind ram pressure projected onto the accretion disk is given by
\begin{equation}
   P_\mathrm{ram, sw}
   =
   \frac{\dot{M}_\mathrm{sw} v_\mathrm{sw}}
        {4 \pi |\vec{r}-\vec{d}|^2}
   |\cos\theta| ,
\label{eq:wind_ram}
\end{equation}
where $\vec{r}$ is the position vector of a point on the accretion disk
measured from the neutron star, $\vec{d}$ is the position vector of the
Be star measured from the neutron star, and $\theta$ is the angle between
the local wind direction and the accretion-disk normal.  Thus,
$|\vec{r}-\vec{d}|$ is the distance from the Be star to the point on the
accretion disk.

\section{Inhibition of accretion by the stellar wind}
\label{sec:criterion}

In this section, we derive an analytical criterion for the wind-driven ablation of accretion disks, adopting a few more assumptions.

First, since in misaligned BeXRB, the accretion disk formation generally occurs around periastron,
we assume that the disk is tidally truncated at $0.4 r_\mathrm{Hill}$ (\cite{Paczynski1977}, \cite{HamiltonBurns1992}; see also \cite{MartinLubow2011}) at periastron, where $r_\mathrm{Hill}$ is the Hill radius approximately given by
\begin{equation}
r_\mathrm{Hill} \simeq a (1-e) \left( \frac{q}{3} \right)^{1/3}
\label{eq:Hill_radius}
\end{equation}
for the binary mass ratio $q \ll 1$, which is the case for BeXRBs. Here, $a$ and $e$ are the semi-major axis and eccentricity of the binary orbit, respectively.

Next, the impact of the wind ram pressure depends not only on $\theta$ but also on the location in the disk. Both depend on the orbital phase and vary from system to system. Considering such large uncertainty and given that $q \ll 1$ means the accretion disk is rather small compared with the periastron separation, we assume that the wind hits the accretion disk with an intermediate angle of 45 degrees above the disk plane ($|\cos \theta| = 1/\sqrt{2}$) and that the distance to the accretion disk is fixed to $a (1-e)$. 
Thus, in our model, the wind ram pressure on the accretion disk is given by
\begin{equation}
   P_\mathrm{ram, sw} \simeq \frac{\dot{M}_\mathrm{sw} v_\mathrm{sw}}{2^{5/2} \pi a^2 (1-e)^2}
\label{eq:wind_ram_approx}
\end{equation}
irrespective of the position on the disk.

\subsection{Ablation of standard accretion disks}
\label{sec:ablation_sd}

Using equation~(\ref{eq:Hill_radius}) and re-normalizing radius by the disk radius, $0.4 r_\mathrm{Hill}$, we rewrite equation~(\ref{eq:pgas_sd}) of the outer disk solution as
\begin{eqnarray}
   P_\mathrm{gas} &=& 6.7 \times 10^2 \alpha_{0.1}^{-9/10} \dot{m}_{16}^{17/20} m_{1.4}^{7/8}
                    \left( \frac{r}{0.4 r_\mathrm{Hill}} \right)^{-21/8} \nonumber\\
                  &\times& q_{0.1}^{-7/8} a_{12}^{-21/8} (1-e)^{-21/8},
\label{eq:pgas_sd2}
\end{eqnarray}
where $a_{12} = a/10^{12}\,\mathrm{cm}$ and $q_{0.1} = q/0.1$.
From equations~(\ref{eq:wind_ram_approx}) and (\ref{eq:pgas_sd2}), we obtain an approximate condition for the wind-driven ablation of standard accretion disks,
\begin{eqnarray}
   \dot{M}_{-9, \mathrm{sw}} v_{8, \mathrm{sw}}
   &>& 1.9 \times 10^3 \alpha_{0.1}^{-9/10} \dot{m}_{16}^{17/20} m_{1.4}^{7/8} 
                    \left( \frac{r}{0.4 r_\mathrm{Hill}} \right)^{-21/8} \nonumber\\
       &\times& q_{0.1}^{-7/8} a_{12}^{-5/8} (1-e)^{-5/8},
\label{eq:wind_vs_sd}
\end{eqnarray}
where $\dot{M}_{-9, \mathrm{sw}}$ and $v_{8, \mathrm{sw}}$ are the wind mass-loss rate normalized by $10^{-9}\,M_\odot\;\mathrm{yr}^{-1}$ and the wind terminal velocity in units of $10^3\,\mathrm{km\;s}^{-1}$, respectively.

This shows that the wind gives different impacts on the accretion disk at different radii. 
We assume, however, that the accretion flow is strongly suppressed if
the wind ram pressure exceeds the disk gas pressure at
$r=0.4r_\mathrm{Hill}$, because the wind can then remove the outer disk
material and prevent efficient inward mass transport.
Then, condition~(\ref{eq:wind_vs_sd}) becomes
\begin{eqnarray}
   \dot{M}_{-9, \mathrm{sw}} v_{8, \mathrm{sw}}
   &>& 1.9 \times 10^3 \alpha_{0.1}^{-9/10} \dot{m}_{16}^{17/20} m_{1.4}^{7/8} \nonumber\\
       &&\times q_{0.1}^{-7/8} a_{12}^{-5/8} (1-e)^{-5/8}
 \label{eq:ablation_sw_sd}
\end{eqnarray}
or for the accretion rate,
\begin{equation}
\dot{M} < \dot{M}_\mathrm{crit},
\label{eq:ablation_cond_sd}
\end{equation}
where the critical accretion rate, $\dot{M}_\mathrm{crit}$, is given by
\begin{eqnarray}
   \dot{M}_\mathrm{crit} &=& 1.4 \times 10^{12} \alpha_{0.1}^{18/17} m_{1.4}^{-35/34} q_{0.1}^{35/34} a_{12}^{25/34} (1-e)^{25/34} \nonumber\\
       && \times \dot{M}_{-9, \mathrm{sw}}^{20/17} v_{8, \mathrm{sw}}^{20/17}\;\mathrm{g\;s}^{-1}.
\label{eq:ablation_mdot_sd}
\end{eqnarray}

\subsection{Ablation of ADAFs}
\label{sec:ablation_adaf}

As in the case of standard disks, we obtain the condition for the wind-driven ablation of ADAFs as
\begin{equation}
\dot{M}_{-9, \mathrm{sw}} v_{8, \mathrm{sw}}
   > 58 \alpha_{0.1}^{-1} \dot{m}_{16} m_{1.4}^{1/2} q_{0.1}^{-5/6} a_{12}^{-1/2} (1-e)^{-1/2} 
\label{eq:ablation_sw_adaf}
\end{equation}
or
\begin{equation}
\dot{M} < \dot{M}_\mathrm{crit},
\label{eq:ablation_cond_adaf}
\end{equation}
with the following critical accretion rate,
\begin{eqnarray}
   \dot{M}_\mathrm{crit} &=& 1.7 \times 10^{14} \alpha_{0.1} m_{1.4}^{-1/2} q_{0.1}^{5/6} a_{12}^{1/2} (1-e)^{1/2} \nonumber\\
       &&\times \dot{M}_{-9, \mathrm{sw}} v_{8, \mathrm{sw}}\;\mathrm{g\;s}^{-1}.
\label{eq:ablation_mdot_adaf}
\end{eqnarray}

\section{Application to individual systems}
\label{sec:indiv}

In this section, we compare the effects of the propeller mechanism and wind-driven ablation for a sample of BeXRBs, comparing the gate accretion rates by the rotating magnetosphere and the critical accretion rates by the wind.
Our sample consists of thirteen systems with well determined or constrained parameters. They are listed in Table~\ref{tab:sample}.

\begin{table*}
\tbl{List of BeXRBs in the Milky Way galaxy to which all parameters to apply our model are known or constrained.}{%
\begin{tabular}{@{}ccc@{\hspace*{3mm}}c@{\hspace*{3mm}}c@{\hspace*{3mm}}c@{\hspace*{3mm}}ccc@{}c@{\hspace*{3mm}}ccc@{}}
\hline
System & Spectral & $M_*$ & $R_*$ & $T_\mathrm{eff}$ & $v_{8, \mathrm{sw}}$$^{(a)}$ & $\dot{M}_{-9, \mathrm{sw}}$$^{(b)}$ & $P_\mathrm{orb}$ & $e$ & $a_{12}(1-e)$$^{(c)}$ & $P_\mathrm{spin}$ & $E_\mathrm{cyc}$ & $B_0$ \\
& type & ($M_\odot$) & ($R_\odot$) & (K) &&& (d) &&& 
(s) & (keV) & ($10^{12}$ G) \\
\hline
4U~0115$+$634 & B0.2V & 19 & 8 & 26,000 & 2.47 & 0.57 & 24.3 & 0.34 & 4.43 & 3.61 & 12 & 1.35 \\
Swift~J0243.6$+$6124 & O9.5V & 18.9 & 7.7 & 32,000 & 2.52 & 5.4 & 28.3 & 0.092 & 6.73 & 9.86 & 146 & 16.4 \\
V~0332$+$53 & O8.5V & 20 & 8.8 & 34,000 & 2.42 & 19 & 34.3 & 0.3 & 6.01 & 4.38 & 28 & 3.15 \\
X~Per & B0V & 17.5 & 7.4 & 30,000 & 2.47 & 2.3 & 250.3 & 0.11 & 27.6 & 837 & 29 & 3.26 \\
A~0535$+$262 & O9.7III & 25 & 15 & 31,500 & 2.07 & 72 & 110.6 & 0.47 & 10.7 & 104 & 50 & 5.63 \\
GRO~J1008$-$57 & B1-2V & 12.6 & 6.1 & 23,200 & 2.31 & 0.075 & 249.48 & 0.68 & 8.95 & 93.5 & 78 & 8.78 \\
1A~1118$-$616 & O9.5V & 18.9 & 7.7 & 32,000 & 2.52 & 5.4 & 24 & $< 0.09$$^{(d)}$ & 6.65 & 407 & 55 & 6.19 \\
GX~304$-$1 & B2V & 10.9 & 5.7 & 20,900 & 2.22 & 0.02 & 132.2 & 0.462 & 9.44 & 275 & 54 & 6.08 \\
2S~1553$-$542 & B1-2V & 12.6 & 6.1 & 23,200 & 2.31 & 0.075 & 31.3 & 0.035 & 6.76 & 9.28 & 23-27 & 2.81$^{(e)}$ \\
Swift~1626.6$-$5156 & B0-2V & 14.2 & 6.5 & 25,500 & 2.37 & 0.26 & 132.9 & 0.08 & 17.5 & 15.36 & 10 & 1.13 \\
KS~1947$+$300 & B0V & 17.5 & 7.4 & 30,000 & 2.45 & 2.5 & 41.5 & 0.034 & 9.01 & 18.7 & 12 & 1.35 \\
EXO~2030$+$375 & B0III & 20 & 15 & 28,000 & 1.85 & 25 & 46.03 & 0.41 & 6.16 & 41.4 & 36$^{(f)}$ & 4.05 \\
& B0V & 17.5 & 7.4 & 30,000 & 2.47 & 2.3 & & & 5.91 & & & \\
SAX~J2103.5$+$4545 & B0V & 17.5 & 7.4 & 30,000 & 2.47 & 2.3 & 12.665 & 0.4055 & 2.52 & 359 & 12$^{(f)}$ & 1.35 \\
\hline
\end{tabular}}
\label{tab:sample}
\begin{tabnote}
$^{(a)}$ $v_{8, \mathrm{sw}} = v_\mathrm{sw}/10^3\,\mathrm{km\;s}^{-1}$. \; 
$^{(b)}$ $\dot{M}_{-9, \mathrm{sw}} = \dot{M}_\mathrm{sw}/10^{-9}\,M_\odot\;\mathrm{yr}^{-1}$. \;
$^{(c)}$ Periastron separation normalized by $10^{12}\,\mathrm{cm}$. \; 
$^{(d)}$ $e=0$ is adopted. \\
$^{(e)}$ $E_\mathrm{cyc}=25\,\mathrm{keV}$ is adopted. \;
$^{(f)}$ Detection claimed but not confirmed.
\end{tabnote}
\end{table*}

\subsection{Gate accretion rates due to rotating magnetospheres}
\label{sec:propeller}

The idea of the classical propeller mechanism is that the magnetosphere of a rapidly rotating neutron star expels accreting gas when the magnetopause rotates at a super-Keplerian speed, effectively inhibiting accretion onto the neutron star \citep{IllarionovSunyaev1975, Stella1986}. The location of the magnetopause, or the magnetospheric radius, $R_\mathrm{m}$, is defined as the radius where the ram or gas pressure of accreting gas is equal to the magnetic pressure of the magnetosphere. 

For the standard-disk case, we adopt the conventional expression
for the magnetospheric radius,
\begin{equation}
R_\mathrm{m, sd} = 4.9\times10^8 k \dot{m}_{16}^{-2/7} m_{1.4}^{-1/7}\mu_{30}^{4/7}\ \mathrm{cm},
\label{eq:rm_sd}
\end{equation}
with $k=0.5$ where $\mu_{30}$ is the magnetic moment of the neutron star, $B_0 R_\mathrm{NS}^3$, normalized by $10^{30}$ (e.g., \cite{Frank2002}). Here, $B_0$ is the surface magnetic field and $R_\mathrm{NS}$ is the radius of the neutron star. In the current analysis, we assume $R_\mathrm{NS} = 10^{6}\,\mathrm{cm}$ for all sample systems.
Equating this radius to the corotation radius, where the rotation frequency of the neutron star is equal to the local Keplerian frequency, i.e.,
\begin{equation}
   R_\mathrm{co} = 1.7 \times 10^8 P_\mathrm{spin}^{2/3} m_{1.4}^{1/3}\;\mathrm{cm},
\label{eq:r_co}
\end{equation}
gives the gate accretion rate, $\dot{M}_\mathrm{gate, sd}$, as
\begin{equation}
\dot{M}_\mathrm{gate, sd} = 3.4 \times10^{16} P_\mathrm{spin}^{-7/3}
m_{1.4}^{-5/3}\mu_{30}^{2}\ \mathrm{g\;s^{-1}},
\label{eq:madotgate_sd}
\end{equation}
below which the magnetic gate closes, because the magnetospheric radius then becomes larger than the corotation radius.

For the ADAF case, instead of adopting the same geometrical factor, we estimate
the magnetospheric radius directly from the pressure balance between the
dipole magnetic pressure and the gas pressure of the self-similar ADAF.
This provides a simple way to account for the geometrically thick nature
of the hot accretion flow.
Using equation~(\ref{eq:pgas_adaf}), we obtain
\begin{equation}
   R_\mathrm{m, adaf} \simeq 2.7 \times 10^8 \alpha_{0.1}^{2/7} 
       \dot{m}_{16}^{-2/7} m_{1.4}^{-1/7} \mu_{30}^{4/7}
       \ \mathrm{cm},
\end{equation}
where $c_1=0.53$ and $c_3=0.35$ have been substituted.
From $R_\mathrm{m, adaf} = R_\mathrm{co}$, we have the gate accretion rate for ADAFs as
\begin{equation}
\dot{M}_\mathrm{gate, adaf} \simeq 5.1 \times 10^{16}
        \alpha_{0.1} P_\mathrm{spin}^{-7/3} m_{1.4}^{-5/3} \mu_{30}^{2}
        \ \mathrm{g\;s^{-1}}.
\label{eq:mdotgate_adaf}
\end{equation}
Thus, the gate accretion rates for the two accretion modes differ only
by a factor of order unity for $\alpha_{0.1}\sim1$, and this difference
does not affect the classification discussed below.

\begin{figure}
\begin{center}
\includegraphics[width=0.8\hsize]{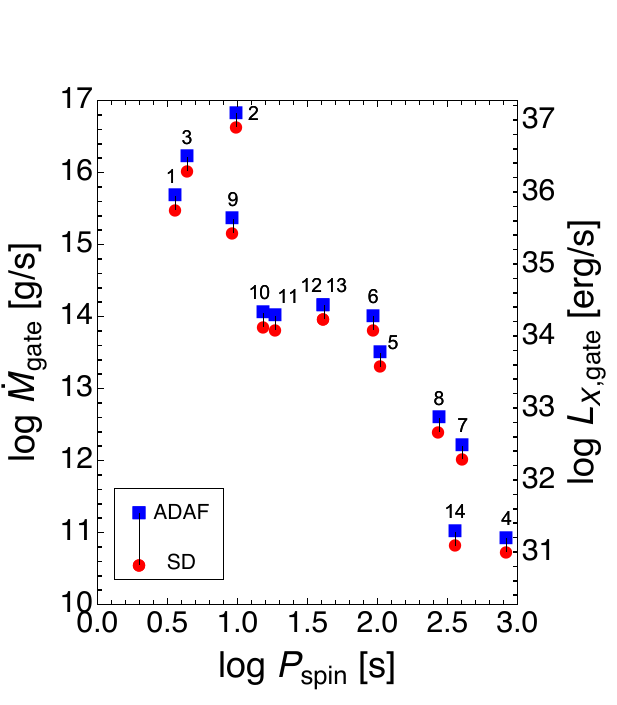}
\end{center}
\caption{Gate accretion rates for the sample BeXRBs. The right axis shows the corresponding X-ray luminosity. They are plotted against the spin period. In each pair of connected symbols, the blue square is for a self-similar ADAF, while the red circle is for a standard accretion disk.
The number attached to each pair of blue square and red circle denotes
1) 4U~0115$+$634, 
2) Swift~J0243.6$+$6124, 
3) V~0332$+$53, 
4) X~Per, 
5) A~0535$+$262, 
6) GRO~J1008$-$57, 
7) 1A~1118$-$616, 
8) GX~304$-$1, 
9) 2S~1553$-$542, 
10) Swift~1626.6$-$5156, 
11) KS~1947$+$300, 
12) EXO~2030$+$375 (B0III), 
13) EXO~2030$+$375 (B0V),
14) SAX~J2103.5$+$4545.
{Alt text: Scatter plot showing that the gate accretion rate decreases systematically with increasing neutron-star spin period for the sample Be X-ray binaries. Each system has two connected points representing the standard-disk and ADAF estimates. The points form a decreasing sequence from high gate accretion rates at short spin periods to low gate accretion rates at long spin periods.}}
\label{fig:gate}
\end{figure}

Figure~\ref{fig:gate} shows the gate accretion rate and the corresponding X-ray luminosity for sample BeXRBs. The latter is calculated by $L_\mathrm{X, gate} = GM_\mathrm{NS}\dot{M}_\mathrm{gate}/R_\mathrm{NS}$. The blue square is for a self-similar ADAF, while the red circle for a standard accretion disk. Each number attached to a pair of symbols denote an individual system. From Figure~\ref{fig:gate}, we note that the gate accretion rate is only weakly dependent on the assumed accretion mode, while it ranges over almost six orders of magnitude. As pointed out by \citet{Okazaki2026}\footnote[2]{Note that equation (3) of \citet{Okazaki2026}, which gives the gate
accretion rate, contains a typographical error: the numerical factor
$1.4\times10^3$ should read $3.8\times10^{17}$.}, the gate X-ray luminosities for short-period systems are compatible with observed transition X-ray luminosities, while those for long-period systems are substantially lower than those to be compatible with observations. It is, thus, unlikely that the propeller mechanism controls the state transition of the latter systems.

\subsection{Mechanism for state transition: The propeller mechanism or wind-driven ablation?}
\label{sec:propeller_vs_wind}

\begin{figure}
\begin{center}
\includegraphics[width=0.8\hsize]{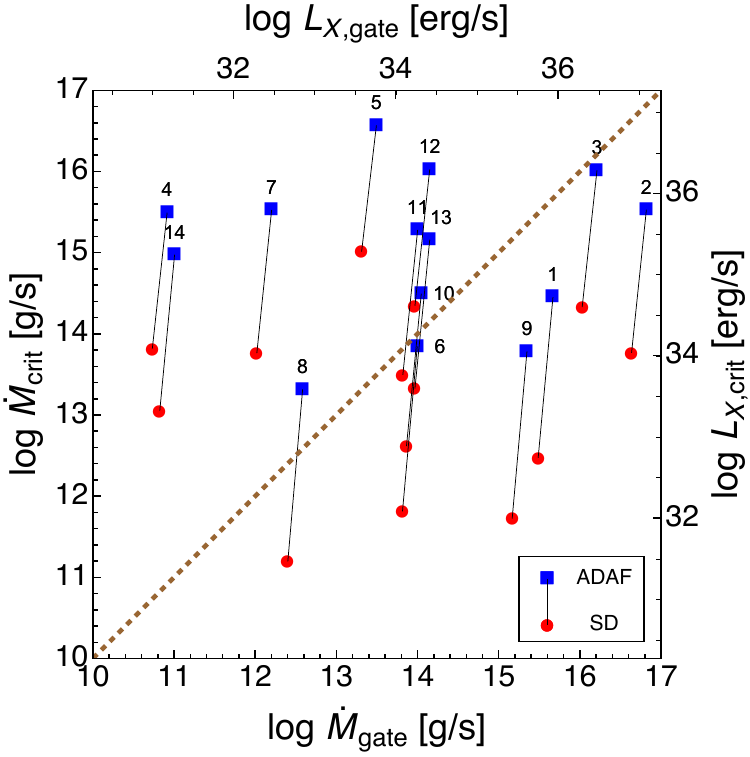}
\end{center}
\caption{Comparison between the critical accretion rate for wind-driven
ablation, $\dot{M}_\mathrm{crit}$, and the gate accretion rate for the
propeller mechanism, $\dot{M}_\mathrm{gate}$, for the sample BeXRBs.
The diagonal dashed line denotes $\dot{M}_\mathrm{crit}=\dot{M}_\mathrm{gate}$.
Systems above this line are expected to be more strongly affected by
wind-driven ablation than by the propeller mechanism, whereas the
opposite is expected for systems below the line.
As in Figure~\ref{fig:gate}, the number attached to each pair of symbols denotes: 
1) 4U~0115$+$634, 
2) Swift~J0243.6$+$6124, 
3) V~0332$+$53, 
4) X~Per, 
5) A~0535$+$262, 
6) GRO~J1008$-$57, 
7) 1A~1118$-$616, 
8) GX~304$-$1, 
9) 2S~1553$-$542, 
10) Swift~1626.6$-$5156, 
11) KS~1947$+$300, 
12) EXO~2030$+$375 (B0III), 
13) EXO~2030$+$375 (B0V),
14) SAX~J2103.5$+$4545.
The upper and right axes show the corresponding X-ray luminosities.
{Alt text: Scatter plot comparing the critical accretion rate for wind-driven ablation with the gate accretion rate for the propeller mechanism. Points above the diagonal line indicate systems where wind-driven ablation is more effective, while points below it indicates systems where the propeller mechanism is more effective.}}
\label{fig:wind_vs_propeller}
\end{figure}

Figure ~\ref{fig:wind_vs_propeller} compares the critical accretion rate, $\dot{M}_\mathrm{crit}$, and corresponding X-ray luminosity, $L_\mathrm{X, crit}$, with the gate accretion rate, $\dot{M}_\mathrm{gate}$, and corresponding X-ray luminosity, $L_\mathrm{X, gate}$. As in Figure~\ref{fig:gate}, the blue and red symbols denote ADAFs and standard disks, respectively, and numbers attached correspond to individual systems. The diagonal dashed line denotes the relation $\dot{M}_\mathrm{crit} = \dot{M}_\mathrm{gate}$. For systems above it, the wind-driven ablation is more important than the propeller to suppress accretion, while the latter is more important than the former for systems below the line. From this figure we find:
\begin{itemize}
\item Four systems (4U~0115$+$634, Swift~J0243.6$+$6124, V~0332$+$53, and 2S~1553$-$542) have $L_\mathrm{X, gate} \gtrsim 10^{35}\,\mathrm{erg\;s}^{-1}$ and $L_\mathrm{X, gate} > L_\mathrm{X, crit}$ irrespective of the accretion mode. In these systems the propeller mechanism is most likely to control the state transition between the outburst and quiescent states.

\item If the accretion mode in all systems is of standard type, A~0535$+$262 is the only system with $L_\mathrm{X, crit} \gtrsim 10^{35}\,\mathrm{erg\;s}^{-1}$ and $L_\mathrm{X, crit} > L_\mathrm{X, gate}$. In this system, wind-driven ablation is likely to contribute to the suppression of
accretion, especially after the accretion rate has declined from the
outburst level. In systems other than five systems mentioned above, neither of the propeller mechanism nor the dynamical ram pressure of the wind is strong enough to cause the state transition.

\item If ADAF is the accretion mode, however, many more systems have $L_\mathrm{X, crit} \gtrsim 10^{35}\,\mathrm{erg\;s}^{-1}$ and $L_\mathrm{X, crit} \gg L_\mathrm{X, gate}$ and form a large group where accretion is inhibited by the stellar wind.
\end{itemize}

\section{Discussion}
\label{sec:discussion}

In this paper, we have investigated the effect of stellar wind on accretion flows in a sample of BeXRBs, assuming they are misaligned systems. In our study, we consider only the dynamical ram-pressure  stripping of an accretion flow at periastron. In this sense, the criterion derived in this paper should be regarded as an
instantaneous strong-ablation criterion.  It tests whether the ram
pressure of the Be-star wind near periastron can exceed the characteristic
pressure of the accretion flow at the disk outer edge.  If this condition
is satisfied, the wind is expected to remove the outer part of the disk
dynamically and inhibit the subsequent accretion.  If it is not satisfied, however, this does not necessarily
mean that the wind is unimportant.  The wind may still reduce the disk
mass through cumulative momentum deposition over many orbital phases,
or may prevent the re-establishment of the disk after the accretion rate
has declined from the outburst level.

Studying the wind effects on two types of accretion flows, standard disks and ADAFs, we have found that ADAFs are much more easily ablated than standard disks.
The large difference between the standard-disk and ADAF estimates on the critical accretion rate is
physically important.  A standard disk has a relatively high mid-plane
pressure at a given accretion rate and is therefore difficult to destroy
instantaneously by the stellar wind.  In contrast, a hot, geometrically
thick, low-density flow is more susceptible to wind stripping.  This
suggests that the wind may not necessarily terminate an outburst-level
standard disk immediately, but may become increasingly important during
the decay phase, when the accretion flow becomes less dense and possibly
more hot-flow-like.

The present study has shown that the propeller mechanism is likely to control the state transition in 4U~0115$+$634, Swift~J0243.6$+$6124, V~0332$+$53, and 2S~1553$-$542. All of these systems have relatively short spin and orbital periods ($P_\mathrm{spin} < 10\;\mathrm{s}$ and $P_\mathrm{orb} \lesssim 30\;\mathrm{d}$).
On the other hand, if the accretion is of standard-disk type, A~0535$+$262, which has relatively long spin and orbital periods ($P_\mathrm{spin} \sim 100\;\mathrm{s}$ and $P_\mathrm{orb} \sim 100\;\mathrm{d}$), is the only system where the wind-driven ablation is a plausible alternative to suppress accretion. However, if the accretion flow is ADAF-like, many systems with long spin periods fall into the same category as A~0535$+$262, for which the propeller mechanism is inefficient.
In this sense, wind-driven ablation and the propeller mechanism may work
as complementary mechanisms for producing X-ray quiescence in BeXRBs.

This interpretation is consistent with our previous SPH simulations \citep{Okazaki2026},
which showed that the Be-star wind can prevent a long-lived accretion
disk from forming in A 0535+26-like systems, while in 4U 0115+63-like
systems the wind mainly weakens the disk and the propeller mechanism
can still dominate the final transition.

\section{Conclusions}
\label{sec:conclusions}

We have studied the effect of the Be-star wind on accretion dynamics in
misaligned Be/X-ray binaries by analytically comparing the wind ram
pressure with the characteristic pressure of the accretion flow.  We have
also compared the resulting critical accretion rate for wind-driven
ablation with the gate accretion rate for the propeller mechanism in a
sample of Galactic BeXRBs.  Our main conclusions are as follows.

\begin{itemize}
\item The propeller mechanism is likely to control the transition to quiescence in short-spin systems such as 4U~0115$+$634 and V~0332$+$53, for which the gate luminosity is comparable to observed transition
luminosities.

\item Wind-driven ablation is a plausible mechanism for suppressing accretion in wide systems with strong winds and slowly rotating neutron stars.
      A~0535$+$262 is the clearest example in the present sample, especially after the accretion rate has declined from the outburst level.

\item The importance of wind-driven ablation depends strongly on the structure of the accretion flow.
      A standard disk is relatively resistant to instantaneous stripping, whereas a hot, low-density ADAF-like flow is much more susceptible to ablation by the stellar wind.

\item The criterion derived here is an instantaneous strong-ablation criterion.
      Longer-term effects, such as cumulative momentum deposition over an orbital cycle and the suppression of disk re-formation, should be examined by hydrodynamical simulations.
\end{itemize}

\begin{ack}
The author used ChatGPT (OpenAI) during the preparation of this
manuscript to improve English wording, clarity, and organization.  The
scientific content, analytical derivations, numerical coefficients,
figures, interpretations, and conclusions were checked and finalized by
the author, who takes full responsibility for the content of the paper.
\end{ack}

\section*{Funding}
 This research was supported in part by JSPS KAKENHI Grant Number 21K03619.



\end{document}